\documentclass[10pt,aps,prl,floatfix,twocolumn,superscriptaddress]{revtex4-1}
\usepackage{graphicx}    % For graph
\usepackage[T1]{fontenc} % Modern font encoding
\usepackage{float}       % For creating charts, graphs and schemes
\usepackage{marginnote}
\usepackage{color}

\begin{document}
%%%%%%%%%%%%%%%%%%%%%%%%%%%%%%%%%%%%%%%%%%%%%%%%%%%%%%%%%%%%%%%%%%%%%
\author{Cono Di Paola}
%\email{cono.di_paola@kcl.ac.uk}
\affiliation{Department of Physics, King's College London, London, WC2R 2LS UK}
\affiliation{Department of Earth Sciences, University College London, London, WC1E 6BT UK}

\author{Roberto D'Agosta}
\email{roberto.dagosta@ehu.es}
\affiliation{Department of Physics, King's College London, London, WC2R 2LS UK}
\affiliation{Nano-bio Spectroscopy Group, Departamento de Fisica de Materiales, UPV/EHU, San Sebastian, 20018 Spain}
\affiliation{IKERBASQUE, Basque Foundation for Science, E-48013, Bilbao, Spain} 

\author{Francesca Baletto}
\email{francesca.baletto@kcl.ac.uk}
\affiliation{Department of Physics, King's College London, London, WC2R 2LS UK}

\title{Topological effects in magnetic platinum nano-particles}
\date{\today}

\begin{abstract}
The magnetic properties of platinum nano-particles ranging in size from a few
to up 300 atoms are investigated through first-principle calculations.
It is found that the total magnetization depends strongly on the local atomic
rearrangements, with an enhancement around five-fold axis. This is due to an elongation of the nearest neighbor distance together with a contraction of the 2$^{nd}$ distance, resulting in a net interatomic partial charge transfer from the atoms lying on the sub-surface layer (donor) towards the vertexes (acceptor). 
\end{abstract}
\maketitle 

The great scientific and technological interest in achieving better magnetic,
optical, and catalytic devices has stimulated intensive research at the
nanoscale, where small and finite size metallic nano-particles (mNP) offer the natural building blocks for a bottom up approach
\cite{sun2000,nishihata2002,koga2004,urban2007,shaw11}. Indeed, for example in
memory storage, a way to overcome
the physical limits in down-scaling the dimensions of the magnetic components
is to design and assemble mNP with controlled geometries in order to maximize the packing density and manipulate their mutual
magnetic interaction, e.g., through doping with natural magnets as Fe and
Co \cite{frey09}. An intriguing possibility is the
building of `super-atoms', i.e., aggregates resembling the behavior of
elemental atoms, even for magnetic purposes as shown by Vanadium and Cesium
NP characterized by a magnetic moment as high as twice that of one
Fe atom in its bulk \cite{reveles09,medel11}. 

A long-standing puzzle of cluster physics is to understand how shape and
 composition influence the physico-chemical properties as a
function of the mNP size: the problem is highly non-trivial because
of the strong dependence of the electronic structure on the interatomic
distances, which are highly distorted especially in non-crystallographic
geometries \cite{baletto05,dipaola2013}. In transition metal nano-systems,
partially occupied and energetically degenerate $d$-orbitals tend to retain to some extent their atomic character \cite{medel11}. This can lead to high
magnetic moments even in non-magnetic bulk species, such as Au, Pd, and Pt due
to a non-uniform distribution of spins in the cluster
\cite{nealon2012,singh2013,shinohara2003,sampedro2003,yamamoto2003,garcia2007,zhang2008,crespo2004,yamamoto2004,medel11,cademartiri11,kowlgi2011,bergeron05,enders2010,kumar08,salazar08,singh08,langerberg2014}.

Pt is a transition metal that shows promising catalytic potential and for this
reason its properties have been investigated at large. In recent
superconducting quantum interference device experiments the magnetic properties
of Pt NP synthesized by wet-chemistry techniques have been observed to depend
strongly on the morphology where branched nano-systems are ferromagnetic
whereas spherical symmetry induces paramagnetism with a blocking temperature
higher than cubic motifs \cite{Liu2006,salazar08}. Furthermore, surfactants are
shown to enhance magnetic moments by breaking symmetry and inducing charge
transfer, raising the total magnetic moment in comparison with amine-coated Pt
nano-materials, where the charge transfer is less efficient \cite{enders2010,
zhang2008}. Deliberately introducing certain geometrical deformations in Pt
nano-wires should be considered as a further fundamental approach devoted to
manipulate and tune the magnetic behavior in Pt nano-systems
\cite{Teng2008,chen2010}. Recent experiments show pronounced
magnetic-conductance in atomic Pt contacts depending on their atomic
configurations \cite{Strigl2015}.

From a theoretical point of view, extensive numerical simulations
based on density functional theory (DFT) have been performed to
understand the magnetic nature of Pt thin films
\cite{blugel1995,niklasson1997} and nano-wires, where the appearance
of ferromagnetism has been predicted only in certain structures
\cite{smogunov1_2008,smogunov2008}. Nonetheless, only a few studies
have been carried out on free Pt NP. In particular, Pt
magnetism in nanoclusters has been associated with the partially
filled $d$ states \cite{Luo2007}. The geometrical distortions
characteristic of a finite nano-object can lead to a shift
between minority and majority spin bands and thus less coordinated
sites may show a higher magnetic moment \cite{dieguez01}. However,
what is the contribution of surface, vertex and core atoms and how an
enhancement of the magnetism in Pt clusters is associated to geometrical factors have never been addressed from an atomistic point of view.

In this Letter, for the first time, the contribution to the total
magnetization arising from different topological environments
--surface-, vertex-, sub-surface- and core-atoms-- in Pt
NP has been calculated, via first-principle techniques.
Different morphologies (icosahedra (Ih), decahedra (Dh), FCC and cubo-octahedra (CO)) for sizes up to 309 atoms have been considered. We find that an important role in understanding their magnetic properties is the
charge transfer between the sub-surface vertex and the vertex. We believe part of this behavior to be generic, i.e.,
it could explain some of the magnetic properties of other
nano-clusters like Au and Pd. Although a ferromagnetic ordering is
always observed in any morphology, the total magnetic moment of
icosahedral structures tops around 150 atoms and it is still
considerable above zero at 309 atoms, while dodecahedral motifs have a
peak in their total magnetization below 100 atoms and then become
gradually non-magnetic. The cubo-octahedra and truncated octahedra show a large
constant magnetization of about 0.2 $\mu_B/atom$ between 30 to 309
atoms, with a dip at 147 atoms.  Quite surprisingly, we find that a
high atomic polarization (AP) is not uniquely associated to atoms with a
low coordination number. An interatomic Heisenberg charge exchange
from sub-vertexes to the vertex determines, indeed,
a large splitting of the majority and minority spin populations
raising even more the AP. This charge transfer has been found to
depend on the local topological environment and enhanced
for vertex and sub-vertex lying on five-fold symmetry axes.

The magnetic behavior of Pt$_N$, where $N$ is the total number of
atoms in the cluster, is calculated using a plane-wave code based on
density functional theory included in the Quantum ESPRESSO package
\cite{QE2009}. Our analysis is not restricted to the \emph{magic} sizes of Ih, Dh and CO shapes, corresponding to 13, 55, 147 and 309 atoms of the above polyhedra, but it includes: uncompleted and poly-icosahedral shapes at 38 and 76 atoms, since they have been found to be promising structural motifs for CoPt systems \cite{parsina12}; the Marks twinning of decahedra at 39 and 75 atoms,
and a FCC motif at 76 atoms which corresponds to an incomplete
octahedron at 85 atoms with a stacking fault and cuts to obtain (100)
facets. All the configurations are ionically relaxed using the
Broyden-Fletcher-Goldfarb-Shann's procedure available within the
Quantum ESPRESSO package. Unrestricted local-spin polarized
calculations are performed, where the exchange-correlation potential
is described self-consistently within the generalized gradient
approximation through the Perdew-Burke-Ernzerhof's functional
\cite{pbe1998}, and the Rabe-Rappe-Kaxiras-Joannopoulos' ultrasoft
pseudo-potential with non-linear core correction is used to model
valence electron-nuclei interactions. The energy cut-off for the plane
wave basis set is put at 45 Ry with a charge density cut-off of 360
Ry. Pt electronic configuration considered is
5d$^9$6s$^1$. Electronic eigenstates are calculated at the $\Gamma$
point only.  A Marzari-Vanderbilt's smearing \cite{marzari1999} with
0.001 Ry as a degauss value is introduced \cite{CMMSE}. Periodic boundary
conditions are applied, and the simulation box includes at least 13
\AA{} of vacuum around the cluster to avoid any spurious effect
between periodic images.
\begin{figure}[t!]
\includegraphics[width=8cm]{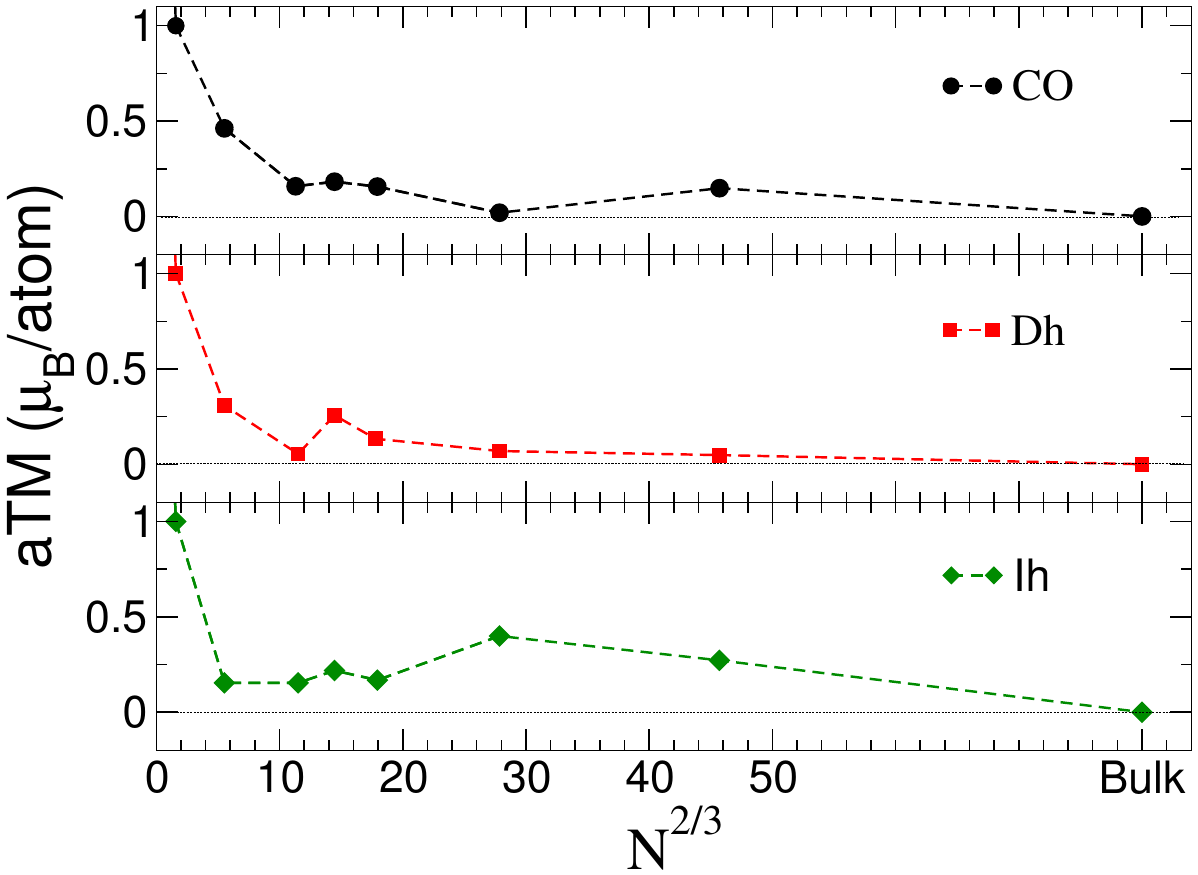}
\caption{(Color online) Atomic averaged magnetization, $aTM$, in $\mu_B/atom$ as
a function of size N$^{2/3}$. The dimer and bulk values are included for comparison. Top panel CO and FCC-like structures; middle panel Dh; bottom Ih.}
\label{fig:TM}
\end{figure}

The macroscopic magnetism of the considered Pt morphologies
is shown in Fig. \ref{fig:TM}, where the atomic averaged total
magnetization ($aTM$), is plotted versus their size. For the
sake of comparison, the dimer and the bulk values are added as
reference points. The $aTM$ is calculated as the total magnetization,
$M_{tot}$, divided by the total number of atoms, $N$, i.e., $aTM=
M_{tot}/N=\int n^{\uparrow}(\textbf{r}) - n^{\downarrow}(\textbf{r})
d\textbf{r}/N,$ where the integral runs over all space, while
$n^{\uparrow}(\textbf{r})$ and $n^{\downarrow}(\textbf{r})$ are the
electron charge densities for spin up and spin down, respectively. For
CO morphology, top panel of Fig. \ref{fig:TM}, the averaged total
magnetic moment presents a flat region at 0.15-0.20 $\mu_B/atom$
between 38 and 309 atoms, with an unexpected dip at 147
atoms. Decahedra peaks in their $aTM$ at 55 atoms with 0.25
$\mu_B/atom$ and then slowly decay to the asymptotic bulk
paramagnetism.  On the other hand, Ih patterns tops of the
averaged total magnetic moment around 150 atoms after an almost
constant behavior at smaller sizes. In the following we are going to
argue that large $aTM$ values are due to an inter-atomic contribution
between sub-vertex atoms lying on the non-crystallographic five-fold
axes and their vertex on top.

The origin of the magnetic properties in mNP can be
rationalized in terms of two contributions: an intra-atomic and an inter-atomic charge transfer. The intra-atomic charge transfer induces an intra-band splitting between majority and minority population and gives rise to
an atomic momentum around the Fermi energy. The inter-atomic charge
transfer mainly occurs between atoms lying on consecutive shells, instead of
intra-shell atoms, and therefore is strongly influenced by the local
topology.
\begin{figure*}[ht!] 
    \includegraphics[width=14cm]{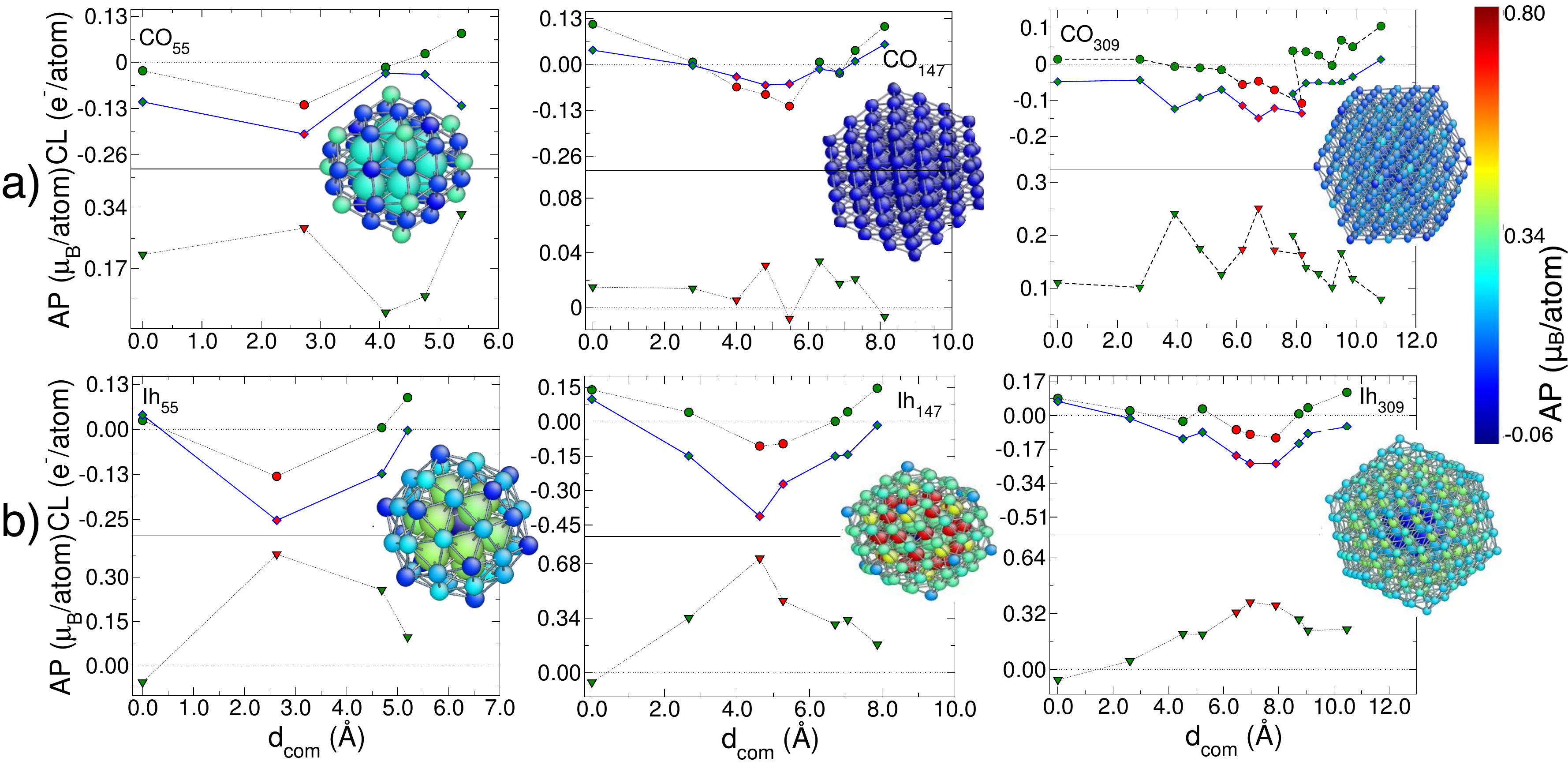} 
    \caption{(Color online)
$AP$ (triangle), total $CL^T$ (circle) and minority $CL^{\downarrow}$ (diamond)
charge losses plotted against the radial distance, $d_{com}$ for a) CO,
and b) Ih shapes at various sizes up to 309 atoms. The symbol color indicates:
red=sub-surface, green=other atoms. In the snapshots, atoms are colored
according to their $AP$ in temperature gradient as shown at the right: (blue) negative for partial charge donors and (red) positive for
acceptors.} 
\label{fig:AP_CL} 
\end{figure*}

To investigate the relationship between the local environment and
magnetic properties, atomic charges are computed performing a Bader
analysis \cite{bader_tang2009,bader1990} over the total as well as
majority and minority spin charge densities. The Bader algorithm,
indeed, searches the 2-D surface on which the charge density between
atoms has a minimum and estimates the charge density inside the volume
found for each atom. This methodology is a very efficient way to
calculate the effective atomic charge. The charge transfer between
shells is estimated in terms of atomic partial charge loss, $CL^T$,
calculated as the difference between the total valence population owned
by an atom and its nominal value of valence electrons mediated over
the equivalent geometrical positions in each shell of a given
morphology. For Pt, the nominal total valence population equals
10. The same approach is adopted to calculate the majority and
minority splitting, the intra-atomic splitting, as a charge loss
$CL^{\uparrow}$ and $CL^{\downarrow}$,respectively, where the nominal
value of valence electrons is now referred to as the up and down
population in the bulk, 5 spins per Pt atom.  The $AP$, or atomic
magnetic moment, is then calculated as the difference between the
majority and minority spin charge losses. The total and minority
charge losses as well as $AP$ are reported in Fig. \ref{fig:AP_CL} as
a function of the radial distance from the centre of mass of the
clusters $d_{com}$, for CO, and Ih, respectively. We want to
point out how sub-vertexes play an important role in the
distribution of local magnetism. Since the smallest size, but where
three shells could be drawn, those atoms are found to show a peculiar
ability to act as donors (negative $CL^T$) of partial charge mainly
towards the low coordinated vertexes (acceptors, positive $CL^T$). The charge donation is generally carried out by the minority population. On the other hand, surface atoms tend to act as acceptors allowing the majority band to
accommodate the charge received. The inter-atomic charge transfer
towards the innermost shells seems to be more complicated but is
substantially still dominated by the sub-surface atomic behavior. Ih, independently of the size, shows a quite
clear behavior both in the AP and CLs, see Fig. \ref{fig:AP_CL}b. The
sub-shell is highly polarized with a net charge transfer towards the
vertexes of the cluster, leading to a significant magnetization in
this morphology. The CO shape instead shows a more varied behavior,
see Fig. \ref{fig:AP_CL}a. For CO$_{55}$ and CO$_{309}$ there is a
peak of AP in the sub-shell and some remnant AP at the vertexes, also
charge is transferred from the core and sub-shells to the surface. At
both sizes there is a similar magnetization per atom. For CO$_{147}$,
on the other hand, the AP is almost vanishing for each shell, and more
significantly the magnetization essentially vanishes.
\begin{figure}[t!]
\includegraphics[width=8cm]{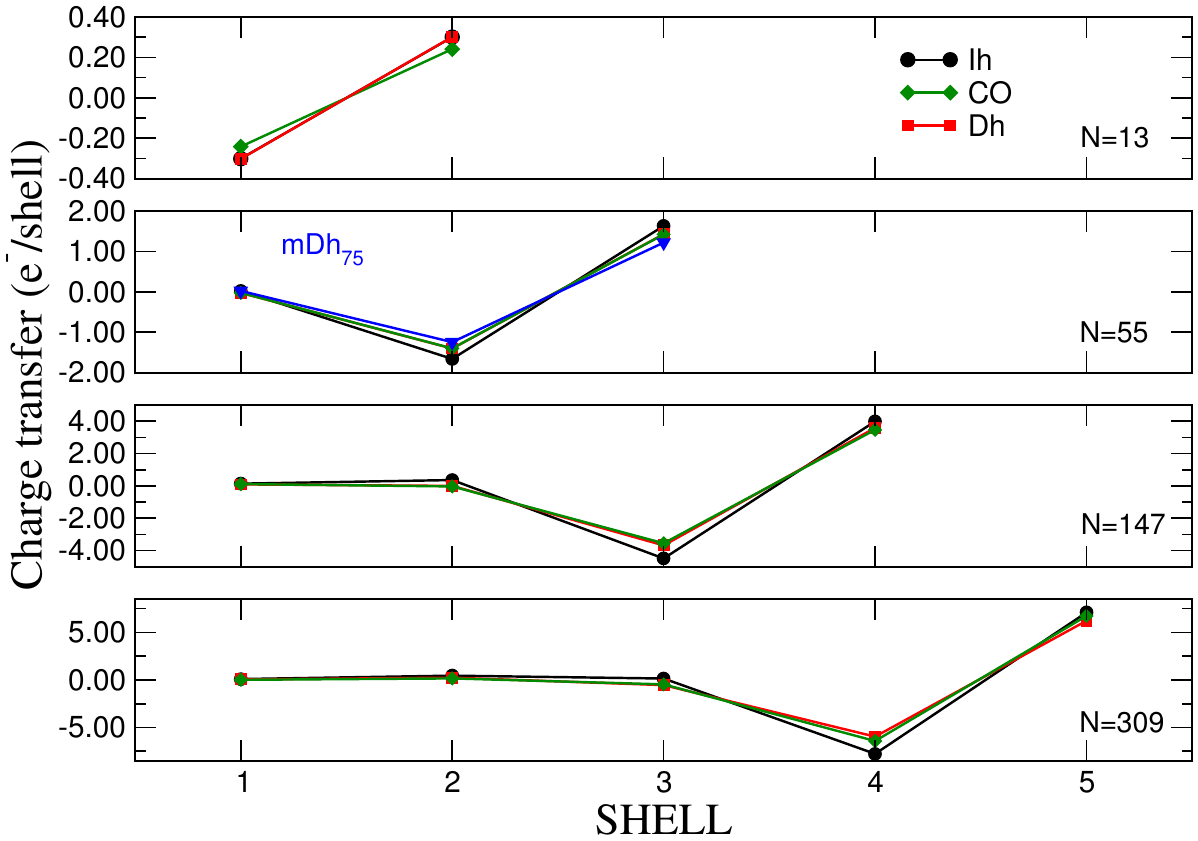}
\caption{(Color online) Total charge transfer (in electron per shell) between neighboring shells. A similar behavior is observed in all the nano-particles considered.}
\label{fig:S2}
\end{figure}

From Fig. \ref{fig:S2}, it is clear that a significant total charge transfer
happens between the sub-surface and the uppermost layer. However, when local
strain is present the distance between vertex and sub-vertex is shorter,
allowing the outermost atoms to act as good charge acceptors. Five-fold
sub-vertexes and vertexes present the shortest distance. The inter-atomic
charge transfer contributes to the local magnetism by acting on the majority
and minority population bringing a specific order to the atomic polarization.
On the other hand, there is another contribution to the magnetic properties
that has to be taken into account, namely the intra-atomic band splitting,
which is dominated by the minority charge loss. The intra-atomic band splitting
is calculated from the total up/down charge loss and the analysis suggested in
the following elucidates different mechanisms related to the local geometrical
arrangements and the size of the NP. Crystalline-like shapes exhibit an initial
intra-atomic charge exchange that decreases with the size of the cluster while
Ih structures are found to show an extraordinary atomic magnetic moment at 147
atoms and a still pretty high TM value at 309 atoms. Dh NP (see Supplemental
Material \footnote{See Supplemental Material for details about the AP and CLs
for Dh NP of size between 55 and 309.}), that contain crystallographic regions
between five-fold axes, behave as a perfect mix up to 55 atoms while two well
separated regions can be considered at bigger sizes: one along (100) direction
characterized by non-magnetic atoms and the other along the 5-fold axis where
the up/down splitting is still effective, therefore an
analysis confined to the first coordination shell is not enough to explain this
separation.

\begin{figure}[t!]
\includegraphics[width=8cm]{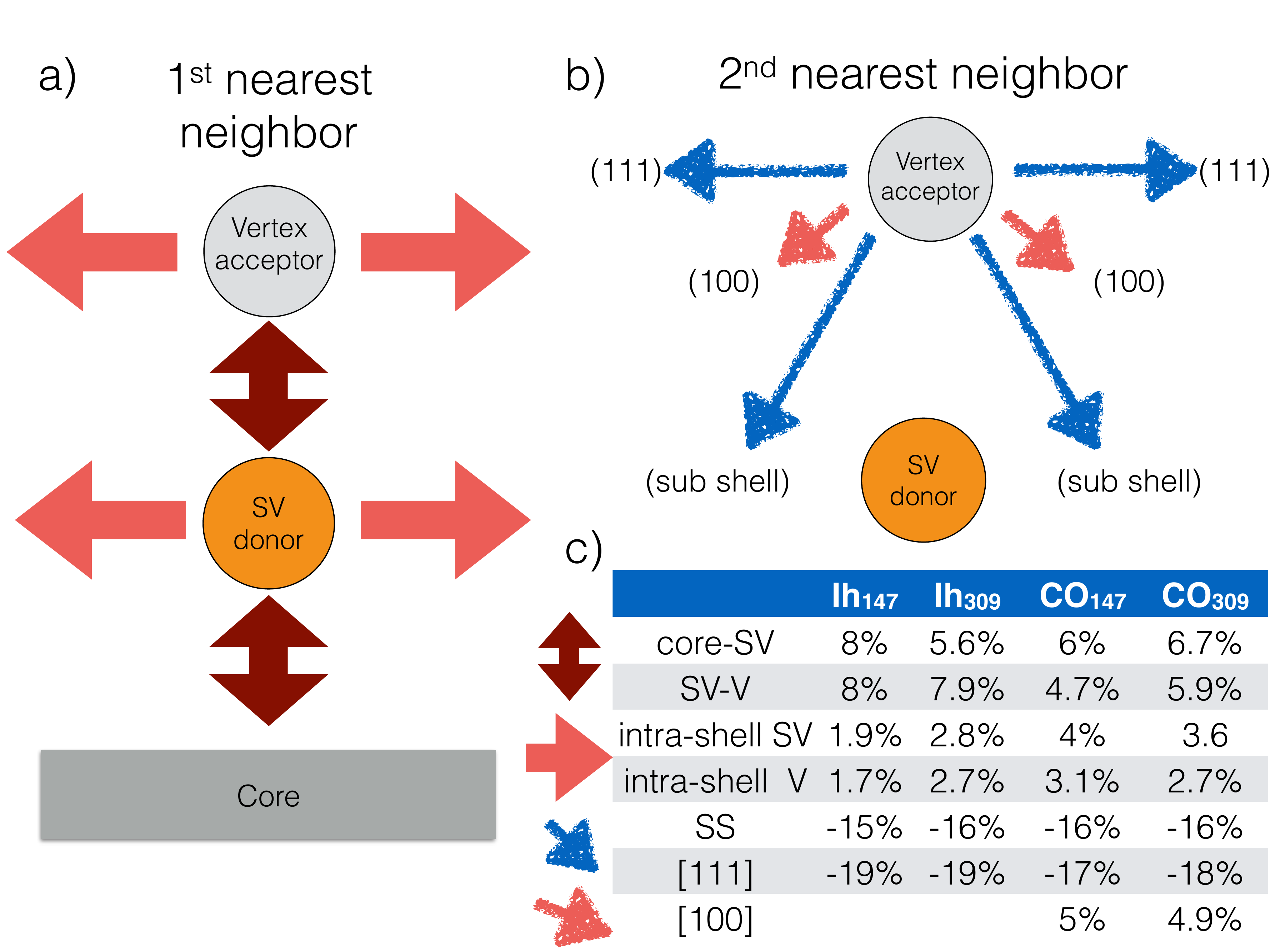}
\caption{(Color online) a) Schematic network of NN average distance of
vertex (V) and sub-vertex (SV). Double arrows stand for inter-shell length while single arrow refers to intra shell distances. b) Schematic 2nd NN network around V: towards (111) surface atoms, common for both Ih and CO; atoms belonging to (100) facets, only for CO. c) Taking the DFT bulk lattice parameter as reference, the Table reports the relative value of the core-SV, SV-V, intra-shell SV, intra-shell V, V-sub shell (SS), (111) 2nd NN, and (100) 2nd NN distances. A positive (negative) value corresponds to a contraction (elongation) with respect to the bulk length.
}
\label{fig:distances}
\end{figure}

Mohn suggested that a further majority/minority band splitting can be justified
by considering the effect of the 2nd nearest neighbor (NN) distance only if
they are enough distorted with respect to their bulk value, where their
contribution is considered negligible \cite{mohn2003}. To understand the
unexpected behavior of the CO$_{147}$ magnetization we have investigated the
distance distribution of first and second nearest neighbor of each vertex and
sub-vertex. In Fig. \ref{fig:distances} we report the vertex and sub-vertex 1st
NN distances; the average distance between vertex and its 2nd NN lying on the
(100) and on (111) surfaces, for CO$_{147}$, CO$_{309}$, Ih$_{147}$, and
Ih$_{309}$. All distances are reported in lattice parameter. These geometries
exhibit a contraction of the 1st NN distance with respect to their bulk value.
However, we observe a stronger compression along the radial direction
core--sub-vertex(SV)--vertex(V), than in the intra-shell directions, see Fig.
\ref{fig:distances} panes a) and c). This seems to justify a charge transfer
from SV towards the core and V, thus generating a finite magnetic moment.
Moreover, the 2nd NN distances between vertex and (111) surface atoms are
always elongated, \ref{fig:distances} b), leading to a reduction of the
intra-shell charge transfer. This effect can further enhance the
majority/minority band splitting. However, in CO, this effect is partially
compensated because the V and (100) atoms 2nd NN distance is slightly
contracted. A further look at the CO$_{147}$ reveals that the intra-shell is
similar to the inter-shell separation. This results in an additional charge
transfer between the atom of either the surface or the 1st sub-shell, then
reducing the effective magnetism as observed in Fig. \ref{fig:AP_CL}.

To summarize, spin-polarized density functional simulations are carried out to
calculate the total magnetization of crystallographic and non-crystallographic
Pt nano-particles for sizes up to 309 atoms. It is observed that an
enhanced magnetism appears in Dh (see Supporting Information) and especially in Ih Pt clusters, while in CO and FCC-like the average magnetization has a almost flat behavior from 36 to 309 atoms (with the notable exception of 147). Through a Bader estimate of the effective atomic charges, we have found that the atomic magnetic properties of Pt$_N$ are ruled not only by the intra-atomic splitting of the up/down bands, but also by the inter-atomic charge transfer between sub-vertexes (donors) and vertexes (acceptors). This contribution is
considerable for atoms lying on a five-fold axis. Our results
demonstrate that the local topology plays a more fundamental
role in explaining the high magnetic moment of Pt nano-particles than
the coordination number itself. In fact, although our results show that a
reduced coordination number can lead to a net atomic magnetization, especially
in CO-like motifs, the sub-vertexes atoms along non-crystallographic axis in
Ih and Dh are responsible of the enhancement of the atomic
polarizability even though their coordination is twelve. This indicates that
the electronic structure and thus the magnetic properties are influenced by the
local environment, in particular associated with the contraction of the
inter-shell distance between the two uppermost layers and the elongation of the second nearest neighbor distances between vertex and (111) surface atoms.

\acknowledgements 
C.D.P. and F.B. have been supported by the U.K. research council EPSRC, under
Grant No. EP/GO03146/1. R.D'A. acknowledges support by DYN-XC-TRANS (Grant No.
FIS2013-43130-P) and NANOTherm (CSD2010-00044) of the Ministerio de Economia y
Competitividad, the Grupo Consolidado UPV/EHU del Gobierno Vasco (Grant No.
IT578-13), and Grant No. MV-2015-1-17 of the Diputacion Foral de Guipuzkoa, and
the hospitality of Sebastian Volz at the Ecole Centrale de Paris for his
hospitality while this work was completed. We thank A. Comisso for useful
suggestions and stimulating discussions, D. Lasa for technical support and
acknowledge the computational time from DIPC.

\bibliography{article}

\end{document}